\documentclass[journal]{IEEEtran}
\usepackage{blindtext, graphicx}

\usepackage{balance}  
\usepackage{fixltx2e}
\usepackage{algorithmicx}
\usepackage{algorithm}
\usepackage[noend]{algpseudocode}
\usepackage{url}
\let\labelindent\relax
\usepackage{enumitem}
\usepackage{multirow}
\usepackage{graphicx}
\usepackage{balance}  
\usepackage{url}
\usepackage[acronym,toc,shortcuts]{glossaries}
\usepackage{epsfig}
\usepackage{amssymb}
\usepackage{amsthm}
\usepackage{multirow}
\usepackage{caption}
\usepackage{subcaption}
\usepackage{siunitx}
\usepackage{soul}
\usepackage{algorithm}
\usepackage{algpseudocode}
\usepackage{csquotes}
\usepackage{bbm}
\usepackage{dsfont}
\usepackage{xcolor}
\usepackage{enumitem}
\usepackage{adjustbox}
\usepackage{array}
\usepackage{multirow}
\usepackage{fancyhdr}
\usepackage[T1]{fontenc}
\usepackage{balance}
\usepackage{comment}

\usepackage[acronym,toc,shortcuts]{glossaries}

\graphicspath{{figures/}}

\newacronym{3GPP}{3GPP}{The 3rd Generation Partnership Project }
\newacronym{5G}{5G}{Fifth Generation}
\newacronym{6G}{6G}{Sixth Generation}
\newacronym{AAA}{AAA}{Authentication, Authorization and Accounting}
\newacronym{AES}{AES}{Advanced Encryption System}
\newacronym{AI}{AI}{Artificial Intelligence}
\newacronym{AP}{AP}{Access Point}
\newacronym{API}{API}{Application Programming Interface}
\newacronym{APN}{APN}{Access Point Name}
\newacronym{AR}{AR}{Augmented Reality}
\newacronym{BS}{BS}{Base Station}
\newacronym{BER}{BER}{Bit Error Rate}
\newacronym{BSSID}{BSSID}{Basic Service Set Identification}
\newacronym{CAT}{CAT}{Capacity-Aware TOPSIS}
\newacronym{CEP}{CEP}{Complex Event Processing}
\newacronym{CELL-ID}{CELL-ID}{cell identification ID}
\newacronym{CGI}{CGI}{Cell Global Identification}
\newacronym{CLSM}{CLSM}{Closed loop spatial multiplexing}
\newacronym{CQI}{CQI}{Channel Quality Indicator}
\newacronym{CN}{CN}{core network}
\newacronym{CNN}{CNN}{Convolutional Neural Networks}
\newacronym{CL}{CL}{Closed-Loop}
\newacronym{CoMP}{CoMP}{coordinated multi-point}
\newacronym{CS}{CS}{central scheduler}
\newacronym{CSI}{CSI}{Channel Status Information}
\newacronym{CPU}{CPU}{Central Processing Unit}
\newacronym{CU}{CU}{Central Unit}
\newacronym{eNB}{eNB}{evolved Node-B}
\newacronym{DL}{DL}{Downlink}
\newacronym{DES}{DES}{Data Encryption Standard}
\newacronym{DMM}{DMM}{Distributed Mobility Management}
\newacronym{DoS}{DoS}{Denial of Service}
\newacronym{DTLS}{DTLS}{Datagram Transport Layer Security}
\newacronym{DU}{DU}{Distributed Unit}
\newacronym{EC}{EC}{Edge Computing}
\newacronym{ECA}{ECA}{Event-Condition-Action}
\newacronym{ECC}{ECC}{Elliptic Curve Cryptography}
\newacronym{eNodeB}{eNodeB}{evolved Node-B}
\newacronym{ECDF}{ECDF}{Empirical Cumulative Distribution Function}
\newacronym{E-RAB}{E-RAB}{E-UTRAN Radio Access Bearer}
\newacronym{ETSI}{ETSI}{European Telecommunications Standards Institute}
\newacronym{FDD}{FDD}{Frequency Division Duplexing}
\newacronym{Fed-KAN}{Fed-KAN}{Federated Learning with Kolmogorov-Arnold Networks}
\newacronym{Fed-MLP}{Fed-MLP}{Federated Learning with Multi-Layer Perceptron}
\newacronym{FEM}{FEM}{Flow Extraction Manager}
\newacronym{FL}{FL}{Federated Learning}
\newacronym{GEO}{GEO}{Geosynchronous Equatorial Orbit}
\newacronym{GGSN}{GGSN}{Gateway GPRS Support Node}
\newacronym{GPRS}{GPRS}{General packet radio service}
\newacronym{GTP}{GTP}{GPRS Tunneling Protocol}
\newacronym{HAPS}{HAPS}{High-Altitude Platform Stations}
\newacronym{HetNet}{HetNet}{heterogeneous network}
\newacronym{HSS}{HSS}{Home Subscriber Station}
\newacronym{HTTP}{HTTP}{Hypertext Transfer Protocol}
\newacronym{HTTPS}{HTTPS}{Hypertext Transfer Protocol Secure}
\newacronym{HDFS}{HDFS}{Hadoop Distributed File System}
\newacronym{HiveQL}{HiveQL}{Hive Query language}
\newacronym{HSPA}{HSPA}{High Speed Packet Access}
\newacronym{IBLER}{IBLER}{Initial Block Error Rate}
\newacronym{ICIC}{ICIC}{inter-cell interference coordination}
\newacronym{ICN}{ICN}{information-centric network}
\newacronym{IEEE}{IEEE}{Institute of Electrical and Electronics Engineers}
\newacronym{IETF}{IETF}{Internet Engineering Task Force}
\newacronym{IMSI}{IMSI}{International Mobile Subscriber Identity}
\newacronym{IMEI}{IMEI}{International Mobile Station Equipment Identity}
\newacronym{IMS}{IMS}{IP Multimedia Subsystem}
\newacronym{ICMP}{ICMP}{Internet Control Message Protocol}
\newacronym{IoT}{IoT}{Internet of Things}
\newacronym{IP}{IP}{Internet Protocol}
\newacronym{IPSec}{IPSec}{Internet Protocol Security}
\newacronym{ISL}{ISL}{Inter-Satellite Link}
\newacronym{ITU}{ITU}{International Telecommunication Union}
\newacronym{IT}{IT}{Information Technology}
\newacronym{GBR}{GBR}{Guaranteed Bit Rate}
\newacronym{GLUE}{GLUE}{General Language Understanding Evaluation}
\newacronym{JSON}{JSON}{JavaScript Object Notation}
\newacronym{KAN}{KAN}{Kolmogorov-Arnold Network}
\newacronym{KPI}{KPI}{Key Performance Indicator}
\newacronym{LA}{LA}{Location Area}
\newacronym{LAC}{LAC}{location area code}
\newacronym{LEO}{LEO}{Low Earth Orbit}
\newacronym{LMA}{LMA}{Local Mobility Anchor}
\newacronym{LSTM}{LSTM}{ Long-Short Term Memory}
\newacronym{LTE}{LTE}{long term evolution}
\newacronym{MADM}{MADM}{Multiple Attribute Decision Making}
\newacronym{MANO}{MANO}{Management and Orchestration}
\newacronym{MCC}{MCC}{Mobile Country Code}
\newacronym{MEC}{MEC}{Mobile Edge Computing}
\newacronym{MCS}{MCS}{Modulation Coding Scheme}
\newacronym{MCP}{MCP}{Management Control Policy}
\newacronym{MNC}{MNC}{Mobile Network Code}
\newacronym{MIMO}{MIMO}{multiple-input multiple-output}
\newacronym{MAG}{MAG}{Mobile Access Gateway}
\newacronym{MAAR}{MAAR}{Mobility Anchor and Access Router}
\newacronym{ML}{ML}{Machine Learning}
\newacronym{MLP}{MLP}{Multi-Layer Perceptron}
\newacronym{MME}{MME}{Mobility Management Entity}
\newacronym{MN}{MN}{Mobile Node}
\newacronym{MNO}{MNO}{Mobile Network Operator}
\newacronym{MSE}{MSE}{Mean Squared Error}
\newacronym{MSISDN}{MSISDN}{Mobile Station International Subscriber Directory Number}
\newacronym{NBI}{NBI}{NorthBound Interface}
\newacronym{NE}{NE}{Network Equipment}
\newacronym{NFV}{NFV}{Network Function Virtualization}
\newacronym{NIST}{NIST}{National Institute of Standards and Technology}
\newacronym{NLP}{NLP}{Natural Language Processing}
\newacronym{NLU}{NLU}{Natural Language Understanding}
\newacronym{NOC}{NOC}{Network Operations Center}
\newacronym{NOMA}{NOMA}{Non-Orthogonal Multiple Access}
\newacronym{NoSQL}{NoSQL}{Not Only SQL}
\newacronym{NR}{NR}{New Radio}
\newacronym{NS}{NS}{Network Service}
\newacronym{NTN}{NTN}{Non-Terrestrial Networks}
\newacronym{QoS}{QoS}{quality-of-service}
\newacronym{QoE}{QoE}{quality-of-experience}
\newacronym{OAM}{OAM}{Operation, Administration and Management}
\newacronym{ONF}{ONF}{Open Networking Foundation}
\newacronym{ONOS}{ONOS}{Open Network Operating System}
\newacronym{OS}{OS}{operating system}
\newacronym{OL}{OL}{Open-Loop}
\newacronym{PDN}{PDN}{packet data network}
\newacronym{PF}{PF}{Proportional Fair}
\newacronym{P-GW}{P-GW}{packet gateway}
\newacronym{PDP}{PDP}{Packet Data Protocol}
\newacronym{PHY}{PHY}{physical layer}
\newacronym{PKI}{PKI}{Public Key Infrastructure}
\newacronym{PMIPv6}{PMIPv6}{Proxy Mobile IPv6}
\newacronym{PMI}{PMI}{Precoding Matrix Index}
\newacronym{PRB}{PRB}{Physical Resource Block}
\newacronym{PUSCH}{PUSCH}{Physical Uplink Shared Channel}
\newacronym{REST}{REST}{Representational State Transfer}
\newacronym{QAM}{QAM}{Quadrature amplitude modulation}
\newacronym{QCI}{QCI}{QoS Class Identifier}
\newacronym{O-RAN}{O-RAN}{Open Radio Access Network}
\newacronym{QRSA}{QRSA}{Quantum Resistant Security Algorithm}
\newacronym{rApp}{rApp}{radio App}
\newacronym{RA}{RA}{Routing Area}
\newacronym{RB}{RB}{Resource Block}
\newacronym{RI}{RI}{Rank Indicator}
\newacronym{RU}{RU}{Remote Unit}
\newacronym{RAN}{RAN}{Radio Access Network}
\newacronym{RFC}{RFC}{Request for Comment}
\newacronym{RIC}{RIC}{RAN Intelligent Controller}
\newacronym{RRC}{RRC}{Radio Resource Control}
\newacronym{RNC}{RNC}{radio network controller}
\newacronym{RNN}{RNN}{Recurrent Neural Networks}
\newacronym{RSSI}{RSSI}{Received Signal Strength Indicator}
\newacronym{RSRP}{RSRP}{Reference Signal Received Power}
\newacronym{RRM}{RRM}{Radio Resource Management}
\newacronym{RT}{RT}{Real Time}
\newacronym{OTT}{OTT}{over-the-top}
\newacronym{SA}{SA}{Stand Alone}
\newacronym{SAC}{SAC}{service area code}
\newacronym{SCMA}{SCMA}{Sparse Code Multiple Access}
\newacronym{SLA}{SLA}{Service Level Agreement }
\newacronym{SDN}{SDN}{Software Defined Networking}
\newacronym{SDO}{SDO}{Standards Developing Organization}
\newacronym{S-GW}{S-GW}{serving gateway}
\newacronym{SINR}{SINR}{signal-to-interference-plus-noise ratio}
\newacronym{SGSN}{SGSN}{Serving GPRS Support Node}
\newacronym{SCO}{SCO}{Service \& Computation Orchestrator}
\newacronym{SFF}{SFF}{Simple-Feed-Forward}
\newacronym{SSID}{SSID}{Service Set Identification}
\newacronym{SVD}{SVD}{singular value decomposition}
\newacronym{TCP}{TCP}{transport control protocol}
\newacronym{TDD}{TDD}{Time Division Duplexing}
\newacronym{TLS}{TLS}{Transport Layer Security}
\newacronym{TM}{TM}{transmission mode}
\newacronym{TN}{TN}{Terrestrial Network}
\newacronym{TEID}{TEID}{tunnel endpoint identifier}
\newacronym{UAS}{UAS}{Unmanned Aerial Systems}
\newacronym{UDN}{UDN}{Ultra Dense Network}
\newacronym{UMTS}{UMTS}{Universal Mobile Telecommunications Service} 
\newacronym{UE}{UE}{user equipment}
\newacronym{UL}{UL}{Uplink}
\newacronym{UDP}{UDP}{User Datagram Protocol}
\newacronym{V2X}{V2X}{Vehicle-to-everything}
\newacronym{VNF}{VNF}{Virtual Network Function}
\newacronym{WiFi}{WiFi}{Wireless Fidelity}
\newacronym{WLAN}{WLAN}{Wireless Local Area Network}

\begin{document}
%
\title{Fed-KAN: Federated Learning with Kolmogorov-Arnold  Networks for Traffic Prediction}


\author{Engin~Zeydan$^{\ast}$, Cristian J. Vaca-Rubio$^{\ast}$,    Luis Blanco$^{\ast}$, Roberto Pereira$^{\ast}$, Marius Caus$^{\ast}$ and Kapal Dev$^{\ddag}$   \\ 
\vspace{.3cm}
$^{\ast}$Centre Tecnològic de Telecomunicacions de Catalunya (CTTC), Castelldefels, Barcelona, Spain, 08860. \\
$^{\ddag}$ CONNECT Centre and Department of Computer Science, Munster Technological University. Ireland. \\
\protect Emails: \{ezeydan, cvaca, lblanco,  rpereira, mcaus\}@cttc.cat, kapal.dev@ieee.org}

\maketitle

\begin{abstract}

Non-Terrestrial Networks (NTNs) are becoming a critical component of modern communication infrastructures, especially with the advent of Low Earth Orbit (LEO) satellite systems. Traditional centralized learning approaches face major challenges in such networks due to high latency, intermittent connectivity and limited bandwidth. Federated Learning (FL) is a promising alternative as it enables decentralized training while maintaining data privacy. However, existing FL models, such as Federated  Learning with Multi-Layer Perceptrons (Fed-MLP), can struggle with high computational complexity and poor adaptability to dynamic NTN environments. This paper provides a detailed analysis for Federated Learning with Kolmogorov-Arnold Networks (Fed-KAN), its implementation and performance improvements over traditional FL models in NTN environments for traffic forecasting. The proposed Fed-KAN is a novel approach that utilises the functional approximation capabilities of KANs in a FL framework.  We evaluate Fed-KAN compared to Fed-MLP on a  traffic dataset of real satellite operator and show a significant reduction in training  and test loss.  
Our results show that Fed-KAN can achieve  a 77.39\% reduction in average test loss  compared to Fed-MLP, highlighting its improved performance and better generalization ability. At the end of the paper, we also discuss some potential applications of Fed-KAN within O-RAN and Fed-KAN usage for split functionalities in NTN architecture. 

\end{abstract}


\IEEEpeerreviewmaketitle


\section{Introduction}

\ac{LEO} satellites play a critical role in bridging connectivity gaps in the 6G era by providing high-speed, low-latency internet in remote areas.  Companies such as Starlink (SpaceX), OneWeb and Telesat are already in the process of launching thousands of satellites into low Earth orbit (300–2000 km altitude) to build a global internet network. However, the increasing demand for seamless and efficient communication in \ac{NTN}, especially in \ac{LEO} satellite systems, also requires advanced strategies to optimize network resources \cite{al2022survey}. Traditional centralized learning approaches struggle to adapt to the dynamic nature of \ac{NTN} due to intermittent connectivity, limited bandwidth and high latency. \ac{FL} has emerged as a promising solution that enables decentralized model training across multiple distributed clients while preserving data privacy and reducing communication overhead. Despite the benefits of \ac{FL}, challenges remain in selecting appropriate machine learning models that can effectively capture the complex and heterogeneous nature of \ac{NTN} traffic. Conventional deep learning models, such as \glspl{MLP}, often require extensive parameter tuning and have difficulty generalizing well under different network conditions. Moreover, the dynamic and time-varying characteristics of NTN traffic require models that can efficiently adapt to non-stationary patterns without incurring excessive computational overhead. 

To address these challenges, we propose  \ac{Fed-KAN} as a novel framework for optimizing network resource allocation in \ac{NTN} environments. \glspl{KAN} offer superior functional approximation capabilities compared to conventional \ac{MLP} \cite{liu2024kan}, by learning more structured and interpretable representations of complex function, making them particularly suitable for capturing complex traffic patterns in satellite-based communication systems.  By integrating KANs into an FL framework as in \cite{zeydan2024f}, we can further enhance predictive accuracy while reducing reliance on frequent satellite-ground communications, thereby improving the overall efficiency of NTN operations. The proposed architecture consists of multiple satellite (beams) acting as \ac{FL} clients, each collecting uplink and downlink traffic data from users within their coverage area. 
The satellites use \ac{KAN}-based local models to predict future traffic patterns and regularly exchange model updates with a central aggregator. The ground station, acting as an \ac{FL} server for aggregation, integrates these updates to refine a global model, which is then distributed back to the to the satellite clients to improve local inference.  

The proposed federated approach not only minimizes bandwidth consumption but also enhances adaptability to spatial and temporal variations in \ac{NTN} traffic patterns. Together with this decentralized approach, adaptive traffic classification and efficient resource allocation for different application categories such as streaming, cloud services, communication and system updates can be enabled efficiently. It can also be used to support intelligent handover mechanisms to optimize the distribution of traffic in overlapping coverage areas of satellite beams. Overall, the integration of \ac{Fed-KAN} mechanism into \ac{NTN} can provide better generalization and interpretability when learning complex traffic behavior and addresses the limitations of traditional FL models in NTN scenarios. In this paper, we provide a comprehensive analysis of performance of \ac{Fed-KAN} compared to traditional FL-based methods such as Federated \ac{MLP}. We demonstrate its effectiveness in predicting network traffic patterns, optimizing resource utilization, and reducing latency in NTN communications.


\subsection{FL and KANs for NTN}

\ac{FL} has gained significant attention in recent years as a distributed machine learning paradigm designed to train models across decentralized clients while preserving data privacy. McMahan et al. \cite{mcmahan2017communication} introduced the foundational concept of \ac{FL}, demonstrating its potential for large-scale collaborative learning without centralizing sensitive data. Subsequent research has explored FL applications in various domains, including edge computing, mobile networks, and satellite communications \cite{duan2023combining}. In the context of \glspl{NTN}, \ac{FL} has been proposed as a solution to mitigate the challenges arising from intermittent connectivity and limited bandwidth in LEO satellite systems. Recent studies \cite{chen2022satellite, jing2022satellite} have explored FL-based architectures for satellite communication networks, focusing on efficient model aggregation and transmission optimization. These approaches aim to reduce latency and improve adaptability to network variations, which is crucial in dynamic \ac{NTN} environments. 

\glspl{KAN}, on the other hand, have emerged as an alternative to traditional neural networks by leveraging learnable activation functions instead of fixed ones \cite{liu2024kan}. This approach enables \glspl{KAN} to achieve superior function approximation, making them particularly suited for complex predictive modeling tasks. Our previous paper \cite{zeydan2024f} introduces a novel \ac{FL} approach that utilises \glspl{KAN} for classification tasks and aims to improve accuracy while preserving privacy. The study shows that \glspl{Fed-KAN} outperform conventional federated \glspl{MLP}  in metrics such as accuracy, precision, recall, F1-score, and stability, suggesting their potential for more efficient and privacy-preserving predictive analytics. Although \glspl{KAN} have been studied mainly in mathematical and scientific contexts, their integration in  \ac{FL} is still largely unexplored in satellite domain. Our work extends this research by proposing a \ac{Fed-KAN} framework specifically tailored to the optimization of \ac{NTN} resources. Furthermore, NTN-based resource allocation has been extensively studied in the literature, with techniques ranging from deep reinforcement learning \cite{naous2023reinforcement} to graph neural networks \cite{tsegaye2024graph}. However, most existing approaches are based on centralized learning architectures that suffer from high communication overhead and a lack of adaptability to the dynamics of real-time traffic. With the introduction of Fed-KAN, we close this gap and offer a decentralized and interpretable solution that increases network efficiency while preserving privacy.

\subsection{Motivation of the paper}

The motivation for the \ac{FL} approach with \ac{KAN} for the traffic prediction problem considered here stems from the unique challenges presented by \ac{LEO} satellites and their dynamic environment. First of all,  \ac{LEO} satellites operate on different schedules, which requires frequent exchange of models to ensure seamless data processing and prediction accuracy for \ac{AI} applications. Due to the constantly changing positions of \ac{LEO} satellites, maintaining consistent communication and synchronization between satellites and ground station is critical when making acurate \ac{AI} decisions. However, traditional centralized learning methods have difficulty adapting to these dynamic schedules. Furthermore, the time schedule for transmission between terrestrial units or \ac{NTN} is highly constrained. The visibility of ground stations is limited, especially for \ac{IoT}-focused satellites, resulting in intermittent connectivity. Therefore, \ac{FL} with the \ac{KAN} approach is particularly advantageous for these scenarios, especially when decentralized model training is required.  \ac{FL} with the \ac{KAN} can provide a robust alternative by enabling local model training and periodic updates of models, reducing the dependency on always on real-time connections with central servers. Our work builds on previous research in \ac{FL} for \ac{NTN} resource optimization, and  application of \glspl{KAN}. We introduce a novel framework that combines these concepts to address the unique challenges of satellite-based communication networks. 



\section{General Architecture}
\label{sec:architecture}

\begin{figure}[htp!]
    \centering
    \includegraphics[width=\linewidth]{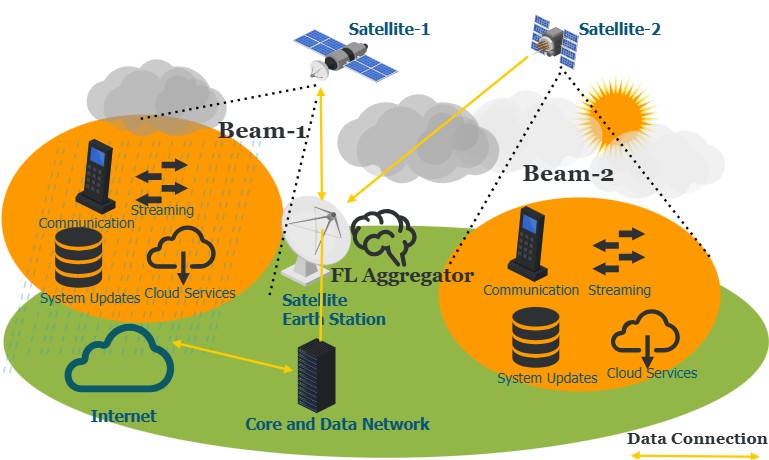}
    \caption{General architecture of NTN and the existence of diverse application traffic on each separate beam of satellites.}
    \label{fig:general_arch}
    \vspace{-.5cm}
\end{figure}

The  \ac{NTN}-based \ac{RAN} consists of \ac{LEO} satellites that are responsible for managing the communication between the users within beams and the ground station connected to data/cloud network \cite{nguyen2024emerging}.  Fig. \ref{fig:general_arch} shows the general architecture in a \ac{NTN}-enabled system, in which several beams enable connectivity to various applications.  Each beam of satellite represent \ac{LEO} satellite beams, each covering a specific group of users within their area of operation.   Each beam also has specific uplink/downlink traffic data from various user applications such as YouTube, Netflix, Instragram, and/or cloud communication services that use the satellite operator's services. The core network and the data network in the ground provide the backend infrastructure for data processing and Internet access. The core network connects the aggregator to the wider data network, enabling large-scale traffic management and predictive analytics.

The architecture of Fig. \ref{fig:general_arch}  can also be used to facilitate decentralized learning and efficient traffic prediction. Multiple beams on each satellite can act as localized \ac{FL} clients that facilitate the transmission of local model updates from the individual beams to the central aggregator (e.g. the ground station), reducing the need for a direct and continuous connection to the core network. The aggregator serves as an \ac{FL} server, aggregating local model updates received from multiple beams and plays a crucial role in classifying and predicting the distribution of traffic.  Once the models are aggregated, the global update is sent back to the beams to refine their local predictions.  A key advantage of the federated framework in this architecture is its ability to enable decentralized learning, ensuring that each beam learns optimally from local traffic patterns. \ac{FL} in this \ac{NTN} scenario can ensure that each beam trains its model independently based on local data, preserving privacy and reducing bandwidth consumption.  This improves the scalability and personalization of the models, while minimizing the communication overhead between the satellites and the ground station and network.  Improved prediction/classification capabilities of \ac{Fed-KAN} can support intelligent resource allocation and optimized handover decisions for \ac{LEO} satellites for better decision making. This is particularly useful for maintaining seamless connectivity in scenarios where satellite beams overlap and enables a smooth handover of traffic flows. 




\section{Federated KAN}
\label{sec:fedKAN}

\subsection{Introduction to Fed-KAN}

\begin{figure}[htp!]
    \centering
    \includegraphics[width=.95\linewidth]{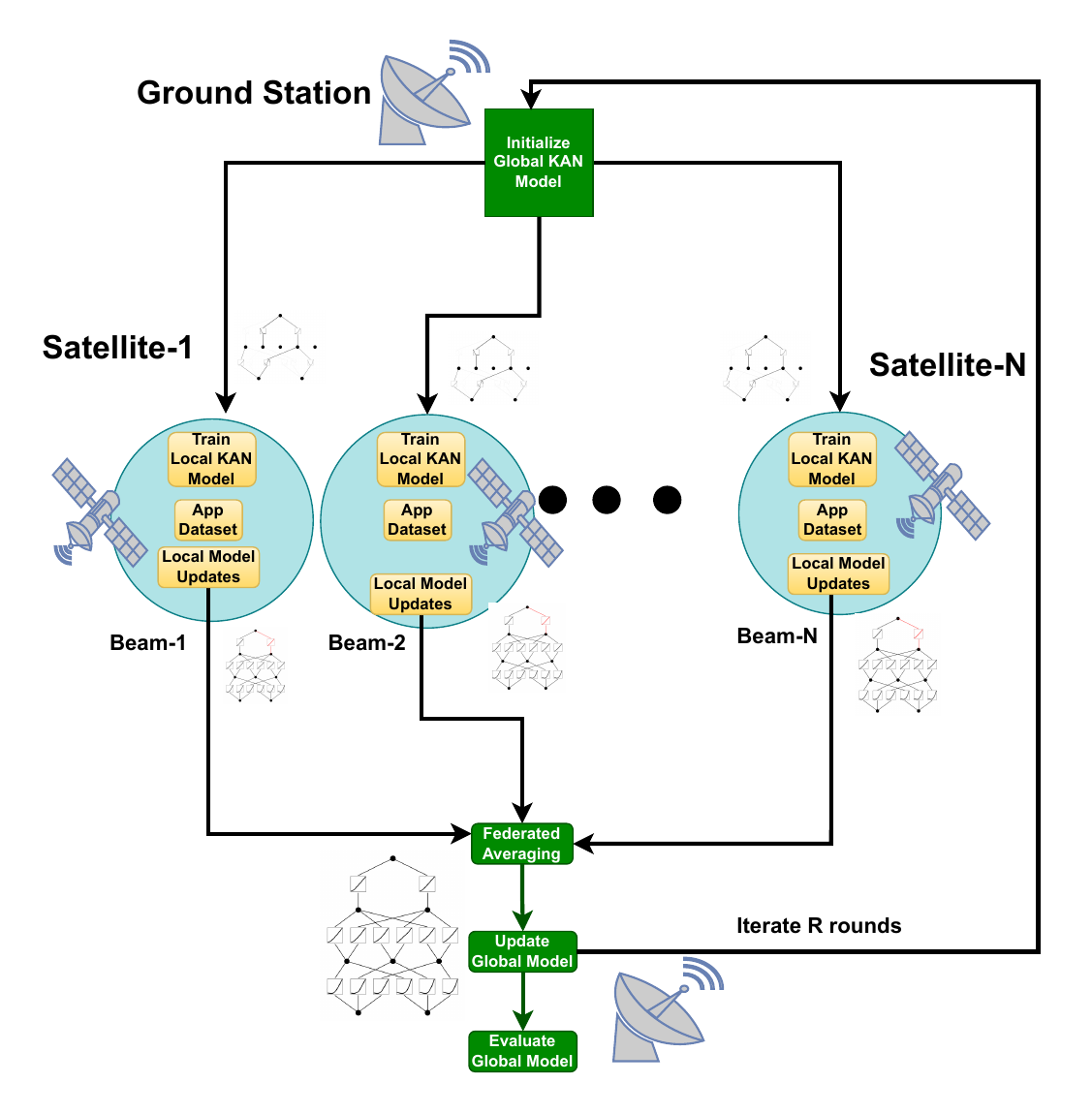}
    \caption{Federated KAN Methodology applied within NTN scenario.}
    \label{fig:Federated_KANs}
    \vspace{-.5cm}
\end{figure}

\glspl{KAN} offer a powerful alternative to traditional neural networks (e.g. \glspl{MLP}) by replacing fixed activation functions with learnable, spline-based transformations. The Kolmogorov-Arnold representation theorem states that any multivariate continuous function can be represented as a composition of simple univariate functions \cite{schmidt2021kolmogorov}. This property makes KANs particularly effective in capturing complex, nonlinear traffic patterns in a distributed environment. Proposed \glspl{Fed-KAN} extend this concept to \ac{FL}, where multiple clients independently train local models while a central server aggregates the model updates without directly accessing raw data.   Fig. \ref{fig:Federated_KANs} shows these steps with nodes representing the main operations and arrows indicating the flow and iteration of the \ac{FL} process for the particular \ac{NTN} scenario where multiple beams in satellites act as  \glspl{Fed-KAN} clients and ground station as  \glspl{Fed-KAN} server aggregator. The \ac{FL} process follows the Federated Averaging (FedAvg) algorithm, where each \ac{Fed-KAN} client trains an independent model on its local data before sending the updated model weights to the central server. The server on the ground aggregates these weights by averaging and iteratively updates the global \ac{Fed-KAN} model over multiple federated rounds. This decentralized learning approach ensures that no raw data is shared between beams, preserving data privacy while taking advantage of collaborative learning across distributed datasets. By leveraging the \ac{KAN} in a \ac{FL} framework, \ac{Fed-KAN} can enable decentralized, privacy-preserving learning for traffic prediction, making it well-suited for satellite network optimization and real-world forecasting applications.

\subsection{Fed-KAN layers}

The architecture of Fed-KANs consists of multiple layers where each \ac{Fed-KAN} layer composed of univariate spline functions. These spline-based transformations allow the network to approximate nonlinear relationships in traffic patterns efficiently. The grid-based representation ensures smooth interpolation between control points, adapting dynamically to fluctuations e.g. in satellite  beam traffic.  The input layer receives two features: Downlink and Uplink traffic values for the last W hours. These inputs are then passed through a series of KAN layers, each responsible for incrementally refining the feature space. The first KAN layer performs feature transformations, while the second KAN layer applies non-linear approximations to capture complex dependencies in the data. The third KAN layer extends this by calculating higher order functions so that the model can learn complicated patterns in network traffic.

In our considered problem with prediction of next hour traffic, after the classical \ac{KAN} transformations, the processed features go through the fully connected (FC) layers that further refine these features, with FC layer 1 containing 8 neurons and FC layer 2 containing 4 neurons, each incorporating ReLU activations and dropout regularization to improve learning stability at each client. Finally, the output layer predicts the percentage traffic values for the next hour for four main traffic categories, namely communication, streaming, cloud services and system updates.


\section{Evaluation Results}
\label{sec:evaluation}

\subsection{Dataset}

The dataset consists of real satellite operator network traffic data collected from four different satellite beams as detailed here \cite{5g_stardust_d4_1}. Each dataset contains time-series traffic measurements, with each row representing a specific timestamp over 743 hours. The dataset is structured into two main categories: input features and target variables. The input features include Downlink and Uplink traffic values. These values indicate the amount of data being transmitted and received over the satellite beams at a given time. The target variables consist of four key categories: Communication, Streaming, Cloud Services, and System Updates.  Table \ref{tab:app_classification} show various applications in each four main classes. The dataset is split in a structured way to ensure effective time series prediction while preserving the integrity of \ac{FL}. For training purposed, first, we construct input-output pairs using a lookback window of $W$ hours. This means that for each prediction, the model uses the data of the last $W$ hours as input to predict the values of the next hour. This creates overlapping sequences to capture temporal dependencies in the data.  Next, we apply an 80-20 split between training and testing, separately for each beam (client), to ensure that 80\% of the data is used for training, while 20\% is reserved for testing. As this is a time series problem, we do not shuffle the data, maintaining the sequential order is essential for accurate future predictions.

\renewcommand{\arraystretch}{1.3}
\begin{table}[h]
    \centering
    \caption{Categorization of applications based on their primary function in the considered dataset.}
    \renewcommand{\arraystretch}{1.3}
    \begin{tabular}{|p{2cm}|p{6cm}|}
        \hline
        \textbf{Category} & \textbf{Applications} \\
        \hline
        Communication & WhatsApp Calls, Telegram, TikTok, Facebook Video, Instagram, VKontakte Video, Google Docs, Office 365 Outlook, Facebook, Reddit, Twitter \\
        \hline
        Streaming & YouTube, YouTube Browsing, Netflix, Quic Obfuscated, Google Services, Amazon Video, Disney+, HBO MAX, Crunchyroll, Acorn TV, Pluto TV, Spotify, TikTok Live, HTTPS Streaming, EA Games, Sony PlayStation Store \\
        \hline
        Cloud Services & Google Cloud Storage, Amazon Cloud, iCloud, OneDrive, Huawei Cloud, MediaFire, Adobe Creative Cloud, Generic CDN, HTTP File Sharing, HTTP File Transfer, Google Photos, Windows Update, Apple Software Update, Steam, Amazon Alexa \\
        \hline
        System Updates & Windows Update, Apple Software Update, Akamai, Generic Web Browsing 3, Advertisements, STUN, BITS, Other UDP, NetEase Games, Epic Games Update \\
        \hline
    \end{tabular}
    \label{tab:app_classification}
\end{table}

\subsection{Model Architecture and Parameters}

\begin{figure}[htp!]
    \centering
    \includegraphics[width=\linewidth]{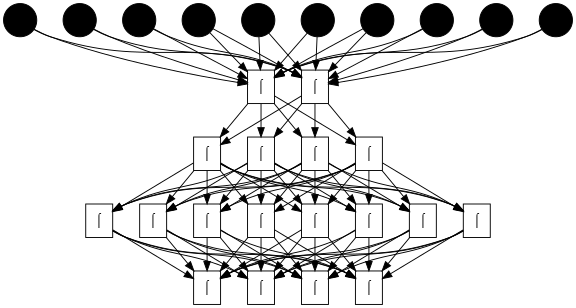}
    \caption{Fed-KAN classifier layers with input (uplink and downlink traffic of last $W$ hours), hidden and output layers (next hour percentage prediction of traffic categories).}
    \label{fig:KAN_Architecture}
\end{figure}

Fig. \ref{fig:KAN_Architecture} shows the architecture of the Fed-KAN classifier in a tree-like structure for a single client. The root node with the black circles is the input layer and each layer applies spline transformations (represented by integral symbols). The edges show how the transformations are connected to each other and form a hierarchical \ac{KAN}. The model \ac{Fed-KAN}, which builds on \glspl{KAN}, was developed to use piecewise polynomials and non-linear functional approximations to model complex relationships in the data.  The input for our \ac{Fed-KAN}  and analyzed \ac{Fed-MLP} models consists of the downlink and uplink traffic of the last $W = 5$ hours, which form a sliding window of historical data. The output is the predicted percentage of different application classes in the next hour.  The Fed-KAN model  features a width configuration of [2, 4, 8] neurons in hidden layers to progressively approximate nonlinear functions. The grid size was set to 5, defining the resolution of the spline transformations, while three Kolmogorov units (k=3) were used for higher-order function learning. A dropout layer with a probability of 0.5 was introduced for regularization before the final fully connected layer, which mapped the transformed features to the output predictions.


The \ac{Fed-MLP} model followed a traditional three-layer architecture with the same hidden sizes to Fed-KAN is used but with [8, 16, 48] neurons to keep the same number of parameters (i.e. 1244). It employed ReLU activation functions in each layer, and dropout (p=0.5) was added to prevent overfitting. The fully connected output layer mapped the learned representations to the final 4-category regression task. Both models were trained using the Adam optimizer with a learning rate of 0.001 and weight decay of 1e-5, ensuring stable convergence. Additionally, gradient clipping (max\_norm=1.0) was applied to prevent gradient explosion. The models were trained over 20 rounds, with each client running 5 local epochs per round using a batch size of 16. For evaluation, \ac{MSE} loss was used as the primary metric, given the regression nature of the task. The global model was validated on each client's test set after every \ac{Fed-KAN} round, with the final performance measured by averaging the test losses across all clients.

\subsection{Implementation details}

The implementation of \ac{Fed-KAN} is structured around four main stages: data preparation, model definition, federated training, and evaluation. The process begins with data preparation, where traffic datasets from multiple satellite beams are preprocessed. For training, input features, in particular uplink and downlink traffic values, are extracted together with target variables, namely communication, streaming, cloud services and system updates. The data is prepared for prediction by applying a time horizon window (e.g. $W=5$ hours) to ensure that past traffic values can be used to predict traffic behaviour for the next 1 hour. After training, the global \ac{Fed-KAN} model is evaluated with test datasets from all beams, calculating the \ac{MSE} loss for performance evaluation.  The model is also compared to a \ac{Fed-MLP} baseline, which features a three-layer fully connected neural network with ReLU activations and dropout. Both models undergo the same federated training process, allowing a direct performance comparison. The trained models are further analyzed through prediction vs. actual value visualizations, providing insights into forecasting accuracy across different traffic categories. 

%


\subsection{Key Observations}

\begin{figure}[htp!]
    \centering
    \includegraphics[width=\linewidth]{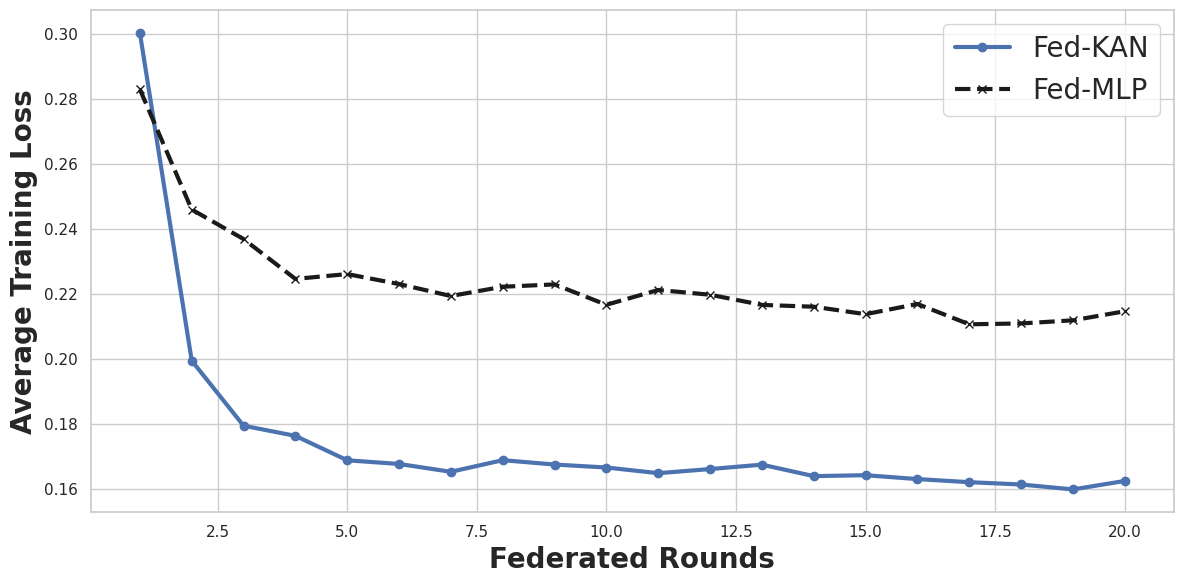}
    \caption{Average Training MSE Loss over Fed-KAN rounds.}
    \label{fig:avg_training_loss}
\end{figure}

\ac{LEO} satellites regularly change satellites to maintain connectivity for users, and reliance on terrestrial gateways can lead to fluctuations depending on gateway utilisation and weather conditions.  Historical network traffic data are used in \ac{Fed-KAN} and Fed-\ac{MLP} models to anticipate future trends in application usage and optimize resource allocation in \glspl{NTN} and \ac{LEO} satellite communication systems.  Both models analyze the past uplink and downlink traffic patterns to predict the distribution of application usage in the next hour. Fig. \ref{fig:avg_training_loss} shows a comparison of the average training losses over 20 federated rounds between \ac{Fed-KAN} and \ac{Fed-MLP} models. The training loss for Fed-KAN decreases sharply in the first rounds and reaches a stable loss value around 0.175 after about 5 rounds, indicating that Fed-KAN converges faster and captures the patterns in the data more efficiently. In contrast, the loss for \ac{Fed-MLP} gradually decreases and stabilizes at a higher loss value of around 0.22. This shows that Fed-KAN outperforms \ac{Fed-MLP} in terms of learning efficiency and achieves a lower training loss across all rounds. The final training loss for Fed-KAN is about 20\% lower than that of \ac{Fed-MLP}, proving its superiority in \ac{FL} scenarios, especially in satellite-based NTN environments where fast adaptation is crucial.

Table \ref{tab:fl_kan_mlp} compares the performance of \ac{Fed-KAN} and \ac{Fed-MLP} based on the average test loss and the number of parameters used during training and testing. The results show that with the same number of parameters, in the test dataset the \ac{Fed-KAN} model outperforms \ac{Fed-MLP} in predicting traffic utilisation by  achieving a 77.39\% reduction in average test loss for predicting the percentages of different application classes using the uplink and dowlink data. This analysis suggests that \ac{Fed-KAN} can be effective in predicting resource utilisation for satellite communication networks where dynamic topology changes and limited computational resources are the main constraints that need to be dynamically addressed.

\begin{table}[htp!]
    \centering
    \caption{Comparison of federated learning models: Fed-KAN vs. Fed-MLP in the considered NTN scenario.}
    \begin{tabular}{|l|c|c|}
        \hline
        \textbf{Model} & \textbf{Number of Parameters} & \textbf{Average Test Loss} \\
        \hline
        Fed-KAN & 1244 & 0.2583 \\ 
        \hline
        Fed-MLP & 1244 & 1.1433 \\ 
        \hline
    \end{tabular}
    \label{tab:fl_kan_mlp}
\end{table}

\section{Fed-KAN Applications in NTNs}

\subsection{Decision Engine and Actuator}

Decision engines and actuators play a crucial role in optimizing the operation of satellite networks. In this phase, there could be three different deployment options that can be used to predict traffic patterns for the next hour over a sliding window period. These options include: (i) Full on-board deployment, where decision making takes place entirely on the satellite. (ii) Partial on-board deployment, where some \glspl{DU} and \glspl{CU} remain on the satellite while others operate remotely. (iv) No on-board deployment, where all functions are processed off-satellite. These flexible deployment strategies allow for optimal resource allocation and reduced latency for network adjustments.  With overlapping satellite beams, an effective handover strategy is critical to maintaining seamless streaming traffic. If an imminent loss of coverage is predicted for a satellite traffic with \ac{Fed-KAN}, the system can preemptively hand over the stream to another satellite or terrestrial station with overlapping coverage, thus ensuring an uninterrupted connection. This proactive approach improves the user experience and optimizes bandwidth usage. This makes \ac{Fed-KAN} an important component of the next generation of satellite communications.

\subsection{Discussions on Split Functions in O-RAN}

Given the distributed and privacy-sensitive nature of \ac{NTN}, \ac{Fed-KAN} can be leveraged as an \ac{AI}-driven intelligence model for optimizing \ac{NTN} operations while preserving data privacy across different \ac{NTN} nodes.  Recently, \ac{O-RAN} has emerged as a new technology that can enable flexible, intelligent, and vendor-neutral deployment of network functions in both terrestrial and \ac{NTN} \cite{campana2023ran, baena2025space}.  An efficient functional split of the \ac{O-RAN} architecture can enable efficient task allocation between the space and ground domains where the proposed \ac{Fed-KAN} algorithm  can be used within this \ac{O-RAN} \glspl{RIC} at various layers. \ac{Fed-KAN} deployment can also be enabled as an xApp or rApp, ensuring dynamic, decentralized, and privacy-preserving \ac{AI}/\ac{ML} model training across the network. The placement of \ac{Fed-KAN} within the \ac{O-RAN} NTN systems may depend on the computing power and communication latency requirements. Note that LEO satellites move fast, requiring constant handover between satellites every 10-15 seconds  as satellites move out of coverage and another takes over the link \cite{fang2024robust}. LEO satellites also have restricted processing capability, making heavy model aggregation infeasible onboard.  Two potential options can be further elaborated.


\subsubsection{Option 1: Near-RT RIC and CU on Ground, DU+RU On-Board (Satellite)} In this scenario-1, the Near-RT RIC and CU can remain on the ground while the DU and RU can be onboard the satellite. The main advantage of this scenario lies in centralized control and reduced computational demand in space. This setup allows for easier software updates and management since the CU and Near-RT RIC remain on ground. In ground-based RIC deployments, \ac{Fed-KAN} can operate as an rApp within the non-RT RIC and perform long-term traffic prediction, adaptive power allocation and beam resource optimization. This allows the central ground station to aggregate model updates from multiple NTN nodes (such as LEO satellites, \ac{GEO} gateways or HAPS) while maintaining privacy through \ac{FL}. However, it leads to latency and intermittent connectivity problems, especially at the F1-C and F1-U interfaces between the CU and the DU and at the E2 interface, which is used for real-time control by the Near-RT RIC. In addition, the O1 management interface may experience interruptions due to the availability of the satellite-ground link. This configuration is ideal when centralized control takes precedence over real-time adaptability and when the satellite's computing resources are limited.

\subsubsection{Option 2: Near-RT RIC and Full gNB On-Board (Satellite)} In this scenario, the complete gNB (CU, DU and RU) is on board the satellite together with the Near-RT RIC, while the Non-RT RIC remains on the ground. This setup reduces latency by eliminating F1 interface communication between the ground and space, enabling faster real-time control and decision making. In addition, \glspl{ISL} can provide more persistent connectivity for satellite-to-satellite communications. However, this approach brings challenges when distributing policies from the Non-RT RIC to the Near-RT RIC via the A1 interface, as these connections may not always be persistent. In addition, O1-based network management from the ground can be more difficult, and the increased computational overhead on the satellite could be a limitation. This option is better suited for missions that require low latency and higher autonomy, but it requires careful consideration of onboard processing capabilities and policy distribution delays between satellite and ground.  The space-based RIC implementation can enable \ac{Fed-KAN} to act as an xApp within the Near-RT RIC, where real-time FL takes place directly onboard satellites, enabling decentralized training across multiple space nodes without relying on a ground link. This deployment is particularly beneficial for real-time beam handover, mobility management and link adaptation in fast-moving LEO constellations.

To summarize, \ac{Fed-KAN}-enabled Non-RT-RIC applications deployed fully on ground can enable  host the aggregator model, execute FL aggregation, and optimize policies based on long-term traffic patterns. However, it can suffer from high latency. In contrast, \ac{Fed-KAN}-enabled near-RT-RIC applications deployed entirely on the satellite model can enable the \ac{FL} process within the satellite network itself by utilising the \glspl{ISL} and on-board computing resources, rather than relying on ground-based server interaction for model aggregation. In this way, model aggregation can be done without frequent ground communication, allowing faster \ac{FL} iterations which can be more suitable fo \ac{LEO} \ac{NTN} systems.  In the meantime, it should be noted that \glspl{ISL} are still not as advanced as ground communication and some satellites still cannot communicate directly with each other and rely on ground stations communications. In such scenarios, hybrid \ac{FL} approaches and configurations can be used, where the satellites can train local \ac{Fed-KAN} models on board and perform partial updates. Later, the aggregation is partially performed on a lead satellite before the updates are sent to the ground-based non-RT RIC for final aggregation. The final global model is later distributed back to the satellites and ground networks.



\section{Conclusions and Future Work}
\label{sec:conclusion}

In this paper, we proposed a \ac{Fed-KAN} enabled  architecture for \ac{LEO} \glspl{NTN} where \ac{KAN} is used within satellites as \ac{FL} client to enable efficient efficient traffic prediction and hence enhanced resource optimization tasks (e.g. resource scheduling) for satellite networks.  The proposed \ac{Fed-KAN} model has exhibited several significant advantages over the conventional \ac{Fed-MLP} models in the context of \ac{NTN} traffic prediction based on real satellite operator data. The use of FL clients (beams) for local model training can improve scalability, data privacy and adaptability.  More specifically, experimental results indicate that \ac{Fed-KAN} achieves a $77.39\%$ reduction in average test loss   compared to \ac{Fed-MLP}, demonstrating improved model  prediction performance. Hence, the capability of \ac{Fed-KAN} to model non-linear relationships in network traffic data and its robustness to fluctuations can improve decision-making in resource allocation and optimization  in satellite connectivity.  Finally, efficient model updates can improve adaptability to real-world operational constraints, ensuring robust and uninterrupted service delivery.  Future work can concentrate on enhancing the capabilities of Fed-KAN for AI-native integrated networks including heterogeneous domains such as non-terrestrial and terrestrial networks (with xHaul, Open RAN, non-public, edge and cloud connectivity). In addition, the integration of the Fed-KAN framework with digital twinning applications  can enable real-time performance and proactive adjustments to improve resilience and service delivery in various domains of integrated networks.  





\section{Acknowledgment}
This work is 
supported by the European Commission under the “5G-STARDUST” Project, which received funding from the SNS JU under the European Union’s Horizon Europe research and innovation programme under Grant Agreement No. 101096573. 

\ifCLASSOPTIONcaptionsoff
  \newpage
\fi

\balance

\bibliographystyle{IEEEtran}
\bibliography{refs}

\begin{thebibliography}{10}
\providecommand{\url}[1]{#1}
\csname url@samestyle\endcsname
\providecommand{\newblock}{\relax}
\providecommand{\bibinfo}[2]{#2}
\providecommand{\BIBentrySTDinterwordspacing}{\spaceskip=0pt\relax}
\providecommand{\BIBentryALTinterwordstretchfactor}{4}
\providecommand{\BIBentryALTinterwordspacing}{\spaceskip=\fontdimen2\font plus
\BIBentryALTinterwordstretchfactor\fontdimen3\font minus \fontdimen4\font\relax}
\providecommand{\BIBforeignlanguage}[2]{{%
\expandafter\ifx\csname l@#1\endcsname\relax
\typeout{** WARNING: IEEEtran.bst: No hyphenation pattern has been}%
\typeout{** loaded for the language `#1'. Using the pattern for}%
\typeout{** the default language instead.}%
\else
\language=\csname l@#1\endcsname
\fi
#2}}
\providecommand{\BIBdecl}{\relax}
\BIBdecl

\bibitem{al2022survey}
H.~Al-Hraishawi, H.~Chougrani, S.~Kisseleff, E.~Lagunas, and S.~Chatzinotas, ``A survey on nongeostationary satellite systems: The communication perspective,'' \emph{IEEE Communications Surveys \& Tutorials}, vol.~25, no.~1, pp. 101--132, 2022.

\bibitem{liu2024kan}
Z.~Liu, Y.~Wang, S.~Vaidya, F.~Ruehle, J.~Halverson, M.~Solja{\v{c}}i{\'c}, T.~Y. Hou, and M.~Tegmark, ``Kan: Kolmogorov-arnold networks,'' \emph{arXiv preprint arXiv:2404.19756}, 2024.

\bibitem{zeydan2024f}
E.~Zeydan, C.~J. Vaca-Rubio, L.~Blanco, R.~Pereira, M.~Caus, and A.~Aydeger, ``F-kans: Federated kolmogorov-arnold networks,'' in \emph{Proceedings of the 1st Workshop on Distributed AI for Enhanced Wireless Networks (DAINET) 2025, in conjunction with IEEE Consumer Communications \& Networking Conference (CCNC)}.\hskip 1em plus 0.5em minus 0.4em\relax Las Vegas, NV, USA: IEEE, January 2025.

\bibitem{mcmahan2017communication}
B.~McMahan, E.~Moore, D.~Ramage, S.~Hampson, and B.~A. y~Arcas, ``Communication-efficient learning of deep networks from decentralized data,'' in \emph{Artificial intelligence and statistics}.\hskip 1em plus 0.5em minus 0.4em\relax PMLR, 2017, pp. 1273--1282.

\bibitem{duan2023combining}
Q.~Duan, J.~Huang, S.~Hu, R.~Deng, Z.~Lu, and S.~Yu, ``Combining federated learning and edge computing toward ubiquitous intelligence in 6g network: Challenges, recent advances, and future directions,'' \emph{IEEE Communications Surveys \& Tutorials}, 2023.

\bibitem{chen2022satellite}
H.~Chen, M.~Xiao, and Z.~Pang, ``Satellite-based computing networks with federated learning,'' \emph{IEEE Wireless Communications}, vol.~29, no.~1, pp. 78--84, 2022.

\bibitem{jing2022satellite}
Y.~Jing, J.~Wang, C.~Jiang, and Y.~Zhan, ``Satellite mec with federated learning: Architectures, technologies and challenges,'' \emph{IEEE Network}, vol.~36, no.~5, pp. 106--112, 2022.

\bibitem{naous2023reinforcement}
T.~Naous, M.~Itani, M.~Awad, and S.~Sharafeddine, ``Reinforcement learning in the sky: A survey on enabling intelligence in ntn-based communications,'' \emph{IEEE Access}, vol.~11, pp. 19\,941--19\,968, 2023.

\bibitem{tsegaye2024graph}
H.~B. Tsegaye and C.~Sacchi, ``Graph neural network-based c-ran monitoring for beyond 5g non-terrestrial networks,'' in \emph{2024 11th International Workshop on Metrology for AeroSpace (MetroAeroSpace)}.\hskip 1em plus 0.5em minus 0.4em\relax IEEE, 2024, pp. 338--343.

\bibitem{nguyen2024emerging}
C.~T. Nguyen, Y.~M. Saputra, N.~Van~Huynh, T.~N. Nguyen, D.~T. Hoang, D.~N. Nguyen, V.-Q. Pham, M.~Voznak, S.~Chatzinotas, and D.-H. Tran, ``Emerging technologies for 6g non-terrestrial-networks: From academia to industrial applications,'' \emph{IEEE Open Journal of the Communications Society}, 2024.

\bibitem{schmidt2021kolmogorov}
J.~Schmidt-Hieber, ``The kolmogorov--arnold representation theorem revisited,'' \emph{Neural networks}, vol. 137, pp. 119--126, 2021.

\bibitem{5g_stardust_d4_1}
\BIBentryALTinterwordspacing
E.~Marín, N.~Trujillo, F.~Parzysz, L.~Reynaud, M.~Caus, L.~Blanco, and C.~J. Vaca-Rubio, ``D4.1: Open data sets for ml-based rrm,'' 5G-STARDUST Project, Tech. Rep., February 2024, accessed: 2025-02-23. [Online]. Available: \url{https://www.5g-stardust.eu/wp-content/uploads/sites/97/2024/09/5G-STARDUST-D4.1-v1.0.F_bis.pdf}
\BIBentrySTDinterwordspacing

\bibitem{campana2023ran}
R.~Campana, C.~Amatetti, and A.~Vanelli-Coralli, ``O-ran based non-terrestrial networks: Trends and challenges,'' in \emph{2023 Joint European Conference on Networks and Communications \& 6G Summit (EuCNC/6G Summit)}.\hskip 1em plus 0.5em minus 0.4em\relax IEEE, 2023, pp. 264--269.

\bibitem{baena2025space}
E.~Baena, P.~Testolina, M.~Polese, D.~Koutsonikolas, J.~Jornet, and T.~Melodia, ``Space-o-ran: Enabling intelligent, open, and interoperable non terrestrial networks in 6g,'' \emph{arXiv preprint arXiv:2502.15936}, 2025.

\bibitem{fang2024robust}
H.~Fang, H.~Zhao, J.~Shi, M.~Zhang, G.~Wu, Y.~C. Chou, F.~Wang, and J.~Liu, ``Robust live streaming over leo satellite constellations: Measurement, analysis, and handover-aware adaptation,'' in \emph{Proceedings of the 32nd ACM International Conference on Multimedia}, 2024, pp. 5958--5966.

\end{thebibliography}
\end{document}